\documentclass[10pt,showpacs,preprintnumbers,amsmath,amssymb,aps,prd,
               nofootinbib,eqsecnum,amsart,
               a4paper]{revtex4}

\usepackage{epsfig}
\usepackage{graphicx,epsf}
\usepackage{color}
\usepackage{bm}
\def\be{\begin{equation}}
\def\ee{\end{equation}}
\def\bea{\begin{eqnarray}}
\def\eea{\end{eqnarray}}
\def\p{\partial}

\def\X{{{\cal{X}}}}
\def\H{{\cal H}}
\def\cs2{c_{\rm{s}}^2}
\def\wt{\widetilde}
\newcommand\eq[1]{Eq.~(\ref{#1})}
\newcommand\eqs[1]{Eqs.~(\ref{#1})}
\newcommand\A[1]{{\phi_{{#1}}}}
\def \eps {\epsilon}
\def \beg {\begin{enumerate}}
\def \en {\end{enumerate}}
\def\M0{{\cal M}_0}
\def\Me {{\cal M}_{\epsilon}}

\def\RR{{\bf{\mathbb{R}}}}
\newcommand\gami[1]{{\gamma_{{#1}}^{~i}}}
\newcommand\gamk[1]{{\gamma_{{#1}}^{~k}}}

\def\fg{{\rm{flat}}}

\begin{document}

\title{A Concise Introduction to Perturbation Theory in Cosmology}
\author{Karim A. Malik$^1$ and David R. Matravers$^2$} 
\affiliation{$^1$
Astronomy Unit, School of Mathematical Sciences, Queen Mary University
of London, Mile End Road, London, E1 4NS, United Kingdom\\
$^2$Institute of Cosmology and Gravitation, University of Portsmouth,
Portsmouth~PO1~2EG,United Kingdom}

\date{\today}
\begin{abstract}
We give a concise, self-contained introduction to perturbation theory
in cosmology at linear and second order, striking a balance between
mathematical rigour and usability. In particular we discuss gauge
issues and the active and passive approach to calculating gauge
transformations. We also construct gauge-invariant variables,
including the second order tensor perturbation on uniform curvature
hypersurfaces.
\end{abstract}

\pacs{98.80.Jk, 98.80.Cq \hfill  
Class.  Quantum  Grav.  25 (2008) 193001,arXiv:0804.3276v2}

\maketitle

\section {Introduction}

Cosmological perturbation theory has recently enjoyed renewed
interest. Linear or first order theory is still a very active field of
research, though the focus has moved on to higher order and even fully
non-linear theory. This is to a large extent due to the availability
of much improved data sets: whereas previously linear theory was
sufficient and the power spectrum the observable of choice, now the
quality and quantity of the data is such, that higher order
observables, as for example the bispectrum, can be compared with the
theoretical predictions.
These new data sets come from observations of the Cosmic Microwave
Background (CMB) on the one hand, such as the one already in progress
by \textit{WMAP} and in the near future also by \textit{PLANCK}. But
also from 21cm surveys on the other hand, mapping the anisotropies in
neutral hydrogen, such as \textit{LOFAR}, now under construction and
\textit{SKA}, currently in its design phase.

Einstein's theory of General Relativity (GR) is highly non-linear, it
is therefore difficult to deal with in all but the simplest situations
using the full theory. Fortunately for cosmologists the universe
appears to the homogeneous and isotropic to a remarkable degree so the
Friedmann-Robertson-Walker (FRW) metric is adequate for many purposes.  For
instance, once known local features are removed, the CMB is isotropic
to an accuracy of $\delta T/T$ of $10^{-5}$. However if we want
greater resolution or more detail then the approximation has to take
into account anisotropy and inhomogeneity. At present this cannot be
done in full generality since we do not have the appropriate exact
solutions to Einstein's equations. This is not surprising, given their
highly non-linear nature. To deal with this problem cosmologists have
resorted to perturbation methods, which have proved effective in other
areas of physics. 
Previous relevant works on perturbation theory in cosmology at a
linear order include
Refs.~\cite{sachs,walker,Bardeen80,KS,MFB,stewart,Ruth}, and at second
order
\cite{Mukhanov96,bruni,MW2003,Noh2004,Nakamura,Bartolo_review}. Beyond
linear order the literature tends to be very technical and difficult
for the non-specialist to follow.
In this paper we aim to strike a balance between mathematical rigour
using the language and tools of differential geometry, and usability
and applicability to the problems of theoretical astrophysics and
cosmology. \\

The essential idea behind perturbation theory is very simple, and best
illustrated by an example for which we choose the metric tensor in
standard cosmology. We assume that we can approximate the full metric
$(g_{\mu \nu})$ of the universe by an expansion
\be
\label{xyz}
g_{\mu \nu} = g^{(0)}_{\mu \nu} + g^{(1)}_{\mu \nu} +\frac{1}{2}
g^{(2)}_{\mu \nu}+\ldots 
\ee 
The metric $g^{(0)}_{\mu \nu}$, called the background, is the FRW
metric with appropriate spatial curvature, i.e. $K = 0,1, -1$
according to the assumptions made about the universe.
The remaining terms are the perturbations of the background. The first
order part is given by
\be
g_{\mu \nu} - g^{(0)}_{\mu \nu} = g^{(1)}_{\mu \nu}\,,
\ee
where the remaining terms are assumed to be negligible compared to
$g^{1}_{\mu \nu}$ and are neglected at {\it first order}. In a similar
way the higher order perturbations can be identified.  This can be
described simply if we assume that the series can be written as
\be
g_{\mu \nu} = g^{(0)}_{\mu \nu} + \epsilon \tilde{g}^{(1)}_{\mu \nu}
+ \epsilon^{2}  \tilde{g}^{(2)}_{\mu \nu} +\ldots
\ee
where the quantities with tildes have absolute magnitudes of order
unity, and we assume that $\epsilon\ll 1$. 
To zeroth order we have $g_{\mu\nu} = g^{(0)}_{\mu \nu}$ and at first
order
\be 
g_{\mu \nu} =
g^{(0)}_{\mu \nu} + \epsilon \tilde{g}^{(1)}_{\mu \nu} \,,
\ee 
and so on using the fact that at each order the higher order terms can
be ignored. In practise it is often a nuisance to introduce the
parameter $\epsilon$ so, where appropriate we will use the form
(\ref{xyz}). Issues of convergence can be removed by working within a
small enough neighbourhood of the background.

Having set up the approximation (\ref{xyz}), we
have to substitute it into the Einstein equations
\be
G_{\mu \nu} + \Lambda g_{\mu \nu} = \kappa T_{\mu \nu}\,,
\ee
to obtain approximate solutions at the required order of approximation
for the application we have in mind. This is more difficult than one
might imagine. Firstly, perturbations of the metric imply
perturbations of the energy momentum tensor, but more importantly,
calculation of the connection coefficients and the Ricci tensor
involves raising and lowering indices and involves terms of different
orders. 
At zero and first order this is not a problem, but at higher orders it
makes the calculations much more complicated and so the choice of
coordinates or form of the metric can be important. Already at second
order we have ``proper'' second order terms and terms quadratic in the
first order quantities.

Another problem arising in cosmological perturbation theory is the
presence of spurious coordinate artefacts or gauge modes in the
calculations. Although GR is covariant, i.e.~manifestly coordinate
choice independent, splitting variables into a background part and a
perturbation is not a covariant procedure, and therefore introduces
this gauge dependence. Prior to 1980 the gauge modes were handled on a
case by case basis, when Bardeen in Ref.~\cite{Bardeen80} resolved the
issue and provided a systematic procedure for eliminating the gauge
freedom at first order. Although it is sometimes argued that the
covariant approach \cite{Ellis:1989jt} avoids the issue of gauge
choice it corresponds to the comoving gauge which is made explicit by
the inclusion of the velocity field \cite{Langlois:2005qp}.  Below we
will address the gauge issue in some detail and explain how it can be
resolved at first and at second order.\\

The paper is organised as follows. In the next section we introduce
perturbation theory using notation and concepts from differential
geometry. In particular we discuss the definition of perturbations and
how perturbations change under small coordinate changes.
In Section \ref{appl_sect} we apply the concepts and results of
Section \ref{pert_sect}. We discuss the construction of
gauge-invariant variables at first and second order. Amongst the
examples discussed is the second order tensor perturbation and how it
can be rendered gauge-invariant.
We discuss our results in Section \ref{dissc_sect}. We finish this
paper with an appendix in which we describe the relevant concepts from
differential geometry used in Section \ref{pert_sect}.

We predominantly use conformal time, $\eta$, related to coordinate
time $t$ by $dt=a d\eta$, where $a$ is the scale factor. Derivatives
with respect to conformal time are denoted by a dash. Greek indices,
$\mu,\nu,\lambda$, run from $0,\ldots 3$, upper case Latin indices,
$A,B,C$ run from $0, \ldots, 4$, while lower case Latin indices,
$i,j,k$, run from $1,\ldots3$.

\section {Perturbation theory}
\label{pert_sect}

In this section we introduce perturbation theory using differential
geometry. Though focusing on cosmology, we keep the discussion
general.  After giving a definition of perturbations we introduce and
define the concept of gauge and study how perturbations change under
gauge transformations.  The relevant definitions from differential
geometry are discussed in Appendix A.

\subsection {Cosmological perturbation theory:
perturbations of space-time}

The application of perturbation methods in space-time brings in a
new problem, since among the physical quantities to be perturbed is the
space-time itself.  Also, because the results will be used in
Relativistic Cosmology, the theory and results must be covariant.
These requirement lead us to base our discussion on the
explicitly coordinate independent
description of Sachs \cite{sachs}, Stewart and Walker
\cite{walker} and Stewart \cite{stewart}. 
In this case by coordinate independent we mean that the
description does not require coordinates and so is intrinsically
covariant. We can and do introduce coordinates to do calculations
and simplify the exposition.
An alternative and widely used, but coordinate dependent, description
is given in Ref.~\cite{MFB}. As one would expect the results are the
same although, in our view, it is easier to understand and see the
source of the final equations in the description of
Ref.~\cite{stewart}. We will however describe both procedures and show
the connection between the two approaches.  \\

\begin{figure}
\begin{center}
\input{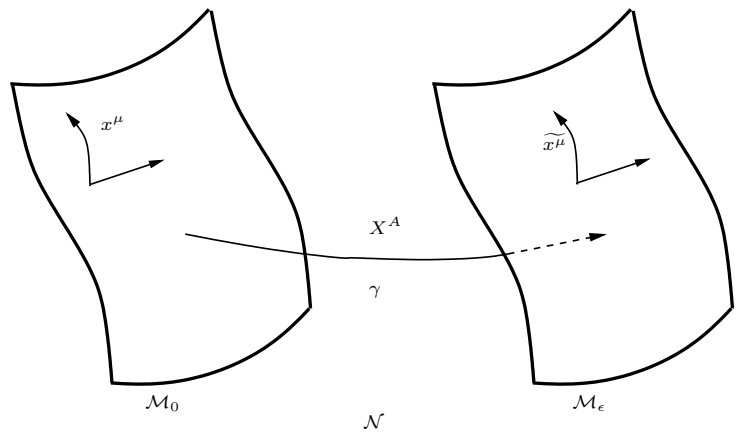}
\caption[pix]{\label{pic1} 
The vector field $X^{A}$ generates a point map between the manifolds
$\M0$ and $\Me$. This in turn yields a diffeomorphism $\phi_{\eps}$
between coordinate neighbourhoods on the manifolds.
 }
\end{center}
\end{figure}

We now follow Stewart \cite{stewart} closely and consider a one
parameter family of 4-manifolds ${\cal M}_{\epsilon}$ embedded in
a 5-manifold ${\cal N}$. Each manifold in the family represents a
perturbed space-time with the base or unperturbed space-time
manifold represented by ${\cal M}_{0}$.  We define a point
identification map $P_{\epsilon}: {\cal M}_{0} \rightarrow {\cal
M}_{\epsilon}$ which identifies points in the unperturbed
manifolds with points in the perturbed manifold. This
correspondence specifies a vector field $X$ upon ${\cal N}$.  This
field is transverse to ${\cal M}_{\epsilon}$ at all points. The
points which lie on the same integral curve $\gamma$ of $X$ are to
be regarded as the same point, see Fig.~\ref{pic1}. This can be
expressed in terms of coordinates. Choose coordinates $x^{\mu}$ on
${\cal M}_{0}$ and extend them to ${\cal N}$ by requiring that
$x^{\mu} = {\rm constant}$ along each of the curves $\gamma$.  This
induces coordinates $\{x^{A} = (x^{\mu}, \epsilon)\}$ with
$A = 0, 1, .., 4$ and $\mu, \nu, .. = 0, 1, .., 3$ on ${\cal N}$.
We parametrise the curves $\gamma$ by $\epsilon$ and so
$dx^{A}/d\epsilon = X^{A}$ and we choose the scaling of
$\eps$ such that
\be 
\phi_{\eps}: {\cal M}_{0}  \rightarrow {\cal M_{\eps}}\,. 
\ee
In this way the vector field $X$ generates a one to one,
invertible, differentiable mapping between ${\cal M}_{0}$ and
${\cal M}_{\eps}$, i.e. a one-parameter group of diffeomorphisms
%
and it follows that $\phi_{\eps}\phi_{\theta} = \phi_{\eps + \theta}$.
In particular the inverse map from ${\cal M}_{\eps}$ to ${\cal
M}_{0}$ will be denoted, in an obvious notation, by
$\phi^{-1}_{\eps} = \phi_{-\eps}$ and the identity map is given by
$\phi_{\eps = 0}$.
%

Given a geometric quantity $T$ defined on ${\cal N}$ the simplest
way to produce a perturbation expansion of $T$ is to expand it as a
Taylor series along $\gamma$.  This yields a covariant power series for
$T$ along the curve.
%
%
To first order the series has the form \cite{stewart}
\be
\phi_{*} T_{\epsilon} 
= T_{0} + \epsilon\left(\pounds_{X}T \right)\Big|_{0} +
O\left(\epsilon^{2}\right)\,,
\ee
%
%
where the $\phi_{*}$ is used to indicate that the quantity is the
pullback, i.e.~it is $T_{\epsilon}$ evaluated at the point where
$\epsilon = 0$.
Lie derivatives are used instead of partial derivatives so that the
series is covariant. For reasons that will become obvious later it is
convenient that the series is pulled back to ${\mathcal M}_{0}$ (see
\eq{A10}).
At higher orders the Taylor expansion is given by \cite{bruni}
\be
\label{phistarpertdef}
\phi_{*}T_{\epsilon} = T_{0} + \sum_{j = 1}^{\infty}
\frac{\epsilon^{j}}{j!}\left(\pounds^{j}_{X} T \right)\Big|_{0}\,,
\ee
where we note again that $\phi_{*}T_{\epsilon}$ is evaluated on ${\cal
M}_{0}$. The expansion automatically provides the covariant
perturbation expansion we want.  Each term in the series is
proportional to a power of $\epsilon$.  The first term $T_{0}$ is
proportional to $\epsilon^{0}$, the background value, the next term
$\epsilon \left(\pounds_{X}T \right)\Big|_{0}$ is proportional to
$\epsilon$ to the first order and so on, and the $n^{th}$ order term is
given by $\frac{\epsilon^{n}}{n!}  \left(\pounds^{n}_{X} T
\right)\Big|_{0}$.

The expansion \eq{phistarpertdef} can be written in a compact and
useful form using the exponential operator,
\be
\phi_{*} T_{\epsilon} = \left(\, e^{(\epsilon \pounds_{X}\,)}\, T
\right)\Big|_{0}\,,
\ee
Here $\phi_{*}T_{\eps}$ is the perturbed value of $T$ pulled back
to ${\cal M}_{0}$ and so the perturbed value of $T$ is given by
\[
\phi_{*} \delta T_{\eps} = \phi_{*} T_{\eps} - T_{0}\,,
\]
where we note that we could not have done the subtraction if we had
not pulled $T_{\eps}$ back to ${\cal M}_{0}$. In an alternative
notation, commonly used in the literature, we include the $\epsilon$
with the $T$ and write
\be
\label{Texp}
T = T_{0} + \delta T
\,,
\ee
where
\be
\label{deltaTexp}
\delta T = T_{1} + \frac{1}{2} T_{2} + \frac{1}{3!}T_{3} + \ldots\,,
\ee
with $ T_{n} = \epsilon^{n}\left(\pounds_{X}^{n}
T\right)_{0}$.\\

In Ref.~\cite{MFB} the approach is as follows. On a single space-time
manifold ${\cal M}$ with coordinates $x^{\mu}$ define a background
model by assigning to all geometric fields $Q$ a fixed background
value $^{(0)}Q$, which is \emph{not} itself a geometric quantity, at each point
on the manifold. While the fields $Q$ may transform as scalar, vector
or tensor fields we require that the $^{(0)}Q$ be fixed functions of the
coordinates.  Under a coordinates transformation the $^{(0)}Q$ will have
the same functional dependence on the new coordinates as they had on
the old ones. A perturbation is then given by
\be
\delta Q = Q - \;^{(0)}Q \,.
\ee

To relate the two approaches we can think of the $^{(0)}Q$ quantity
playing the role of a quantity defined on ${\cal M}_{0}$ in the
Stewart description and the coordinate change corresponding to a
change of coordinates on ${\cal M}_{\eps}$. But it is important to
note that the approach of Ref.~\cite{MFB} only one manifold is
necessary.
The Stewart approach \cite{stewart} avoids the need for the quantity
$^{(0)}Q$, which is not covariant 
and gives a simple diagrammatic representation at the price of having
to introduce the abstract 5-dimensional manifold ${\cal N}$.
However, note that it is the split into a background and a
perturbation which in general is not covariant. This split is common
to both approaches and it gives rise to the gauge dependence.
%

\subsection {Gauge Transformations}
\label{gaugetrans_sect}

Gauge is arguably the most over-used word in mathematics and physics.
Sometimes the meanings are related but often they are not and it is a
waste of time trying to relate them. To avoid confusion we recommend
that the word ``gauge'' as used here is interpreted as defined and not
related to other uses of the word. The choice of correspondence
between points on ${\cal M}_{0}$ with those on ${\cal M}_{\epsilon}$
or, equivalently, the choice of a vector field $X$ is a gauge
choice. The vector field $X$ is called the generator of the gauge.

Let us now turn to defining gauge dependence in a clearer
way. Consider a point $p$ in ${\cal M}_{0}$ and the generators $X$ and
$Y$ corresponding to two different gauge choices (see
Fig.~\ref{pic2}). The choice $X$ will identify point $p$ on ${\cal
M}_{0}$ with a point $q$ on ${\cal M}_{\epsilon}$ and will assign to
$q$ the same $x^{\mu}$ coordinates as at point $p$. On the other hand
the gauge choice $Y$ will identify $p$ with a different point $u$ on
${\cal M}_{\epsilon}$ assigning in its turn the coordinates of $p$ to
$u$. Clearly the choice of gauge induces a coordinate change (a gauge
transformation) on ${\cal M}_{\epsilon}$. This interpretation is
called the \emph{passive view} Ref.~\cite{MFB,bruni}. \\

\begin{figure}[h]
\hfill
\begin{minipage}[t]{.45\textwidth}
\begin{center}  
\input{fig2.pstex_t}
\end{center}
\end{minipage}
\hfill
\begin{minipage}[t]{.45\textwidth}
\begin{center}  
\input{fig3v2.pstex_t}
\end{center}
\end{minipage}
\hfill
\caption{ On the left panel, the \emph{passive view}: The point $p$ on the
manifold $\M0$ is mapped to two different points $q$ and $u$ on $\Me$
depending on the choice of gauge, corresponding to the choice of
vector field, we make.
On the right panel, the \emph{active view}: the points $p$ and $q$ on
$\M0$ both map to the point $u$ on $\Me$. Again the choice of gauge
determines the mapping.  The vector fields generate the gauge
choice. A change in gauge from $X^{A}$ to $Y^{A}$ produces a gauge
transformation.
} 
\label{pic2}
\end{figure}

\begin{figure}
\begin{center}
\input{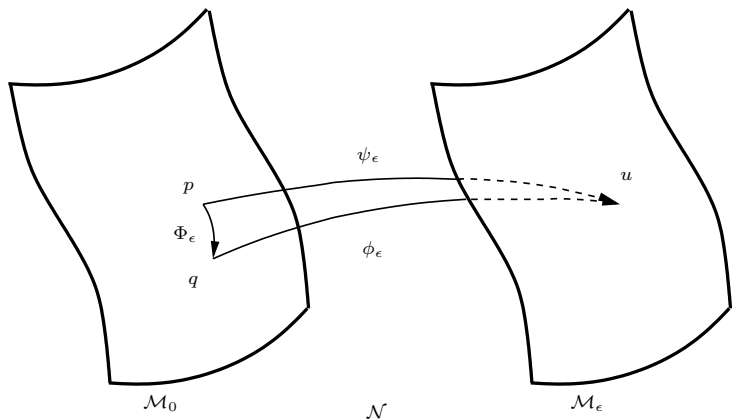}
\caption[pix]{
\label{pic3alt}
For the active point of view of a gauge transformation we choose a
point on $\Me$ and determine the points $p$ and $q$ on $\M0$ which map
to $u$ under the gauge choices $\psi_{\eps}$ and $\phi_{\eps}$.
The map $\Phi_{\eps}$ which maps the point $p$ to the point $q$ is
then formed by first mapping $p$ to $u$ using the map $\psi_{\eps}$
and then mapping $u$ to $q$ using the map $\phi_{-\eps}$.  Thus
$\Phi_{\eps} = \phi_{-\eps}\circ \psi_{\eps}$. 
}
\end{center}
\end{figure}

For the \emph{active view} we choose a point $u$ on ${\cal M}_{\epsilon}$
and find the point $p$ on ${\cal M}_{0}$ which maps to $u$ under
the gauge choice $X$ and the point $q$, also on ${\cal M}_{0}$,
which maps to $u$ under the gauge choice $Y$, see Fig.~\ref{pic3alt}.
The gauge transformation this time is defined on ${\cal M}_{0}$
and takes the coordinates of $q$ to those of $p$ in one of the two
choices of gauge. 

In summary as we shall explain in more detail below, in the active
approach the transformation of the perturbed quantities is evaluated
at the same coordinate point, whereas in the passive approach the
transformation is taken at the same physical point.\\

In the passive approach of Ref.~\cite{MFB} the role of the background
manifold is played by the background quantities $^{(0)}Q$ and the
coordinate transformation corresponding to the gauge choice only
affects the geometric quantities $Q$. The perturbation is the
difference between $^{(0)}Q$ and $Q$ so only half the quantities
determining the perturbation are transformed by the gauge
transformation.\\

The gauge dependence in perturbation theory stems from the fact that
we separate quantities into a background and a perturbed part, a
operation not covariant in general, which introduces additional,
unphysical degrees of freedom. However, as shown below in Section
\ref{appl_sect}, by choosing and combining suitable matter and metric
variables the gauge dependencies can be made to cancel out (the
quantities so constructed will not change under a gauge
transformation). This process is equivalent to choosing suitable
physical hypersurfaces, say comoving or of uniform curvature
\footnote{One should not confuse gauge independence in perturbation
theory with what is called gauge choice in general relativity which
arises from coordinate invariance: the Bianchi identities introduce
four additional degrees of freedom into the Einstein equations. This
allows to always choose four free functions in the metric, i.e.~four
particular coordinate functions, that might simplify the problem under
consideration if suitably chosen.}.

\subsubsection{Active point of view}
\label{active_subsect}

To take the argument further we will now focus on the active
interpretation of the gauge transformation. Corresponding to the
gauge choice $X$, i.e.~the choice of the vector field $X$ transverse
to ${\cal M}_{0}$ we have a diffeomorphism $\phi_{\epsilon}$ where
$\phi_{\epsilon} : {\cal M}_{0} \rightarrow {\cal M}_{\epsilon}$ and
corresponding to the vector field $Y$ we have a diffeomorphism
$\psi_{\epsilon}: {\cal M}_{0} \rightarrow {\cal M}_{\epsilon}$. For
all $\epsilon$ these two vector fields induce a diffeomorphism (gauge
transformation) $\Phi_{\epsilon}$ on ${\cal M}_{0}$ given by, see
Fig.~\ref{pic3alt},
\be
\Phi_{\epsilon} : {\cal M}_{0}  \rightarrow  {\cal M}_{0}\,,
\ee
where $\Phi_{\eps}$ is made up of two parts - a map $\psi_{\eps}$ from
${\mathcal M}_{0}$ to ${\mathcal M}_{\eps}$ and a map $\phi_{-\eps}$
from ${\mathcal M}_{\eps}$ to ${\mathcal M}_{0}$, i.e.
\be
\Phi_{\epsilon} : = \phi_{-\epsilon} \circ \psi_{\epsilon}\,.
\ee
Under this gauge transformation the transformation of a geometric
quantity $T$ is given by (see appendix \eq{A6} and \eq{A9})
\begin{eqnarray}
\Phi_{* \epsilon} T & = & \left(\phi_{-\epsilon} \circ
\psi_{\epsilon}\right)_{*} \,  T \\
& = & \psi_{* \epsilon} \circ \phi_{* -\epsilon} \, T \\
& = & e^{(\epsilon \pounds_{\psi_{X}})} \,
e^{(- \epsilon \pounds_{\phi_{Y}})} \, T\,,
\end{eqnarray}
where we have used the fact that the pull-backs of the
transformations induced by the gauge choices can be written as
Taylor series in terms of the exponential notation as
\begin{eqnarray}
\phi_{* \eps} T & = & e^{\eps \pounds_{\phi_{X}}} T\,, \\
\psi_{* \eps} T & = & e^{\eps \pounds_{\psi_{X}}} T\,,
\end{eqnarray}
(for more details again see the Appendix \eq{A6} to \eq{A9}).  Also note
that the $T$ here have to be evaluated on ${\cal M}_{0}$ but putting
$T_{0}$ would be confusing.\\

Now we invoke the Baker-Campbell-Haussdorf formula \cite{sopuerta}
which enables us to write $\Phi_{*\epsilon} T$ in the following form
\be
\label{17a}
\Phi_{* \epsilon} T = \exp \left(\sum_{n = 1}^{\infty}
\frac{\epsilon^{n}}{n!}\pounds_{\xi_{n}}\right) T
\,,
\ee
where
$$
\xi_{1} = Y - X, \ \ \ \xi_{2} = [X, Y], \ \ \ {\rm and} \ \ \
\xi_{3} = \frac{1}{2}[X + Y, [X, Y]]\,,
$$
and $X$ and $Y$ are the gauge generators, i.e. the vectors which
determine the gauge choices. Explicitly the first few terms of the
gauge transformation \eq{17a} are
\begin{eqnarray}
\label{2.16}
\Phi_{* \epsilon} T & = & T\big|_{0} + \eps \pounds_{\xi_{1}} T\big|_{0} +
\frac{\eps^{2}}{2} (\pounds_{\xi_{2}} +
\pounds^{2}_{\xi_{1}})) T\big|_{0} \nonumber \\
& & \mbox{} +  \frac{\eps^{3}}{3!} \left( \pounds_{\xi_{3}} +
\frac{3}{2} \left[\pounds_{\xi_{1}}, \pounds_{\xi_{2}}\right] +
\pounds^{3}_{\xi_{1}} \right)T\big|_{0} + O(\eps^{4}))\,,
\end{eqnarray}
where we indicate that $T$ has to be evaluated on the manifold $\M0$
by the notation $T\big|_{0}$.

If we now use the equation (\ref{Texp}) to introduce the results
of Taylor expanding $T$ into the formula \eq{2.16} we obtain
\begin{eqnarray}
\label{2.18}
\tilde{{\bf T}}_{0} & = & {\bf T}_{0}\,,\nonumber \\
\tilde{{\bf T}}_{1} & = & {\bf T}_{1}
+ \pounds_{\xi{1}}{\bf T}_{0} \,,\nonumber \\
\tilde{{\bf T}}_{2} & = & {\bf T}_{2}
+ \pounds_{\xi_{2}}{\bf T}_{0}
+  \pounds_{\xi_{1}}^{2}{\bf T}_{0}
+ 2 \pounds_{\xi_{1}}{\bf T}_{1}\,,
\end{eqnarray}
where $\xi^\lambda$ is the vector field generating the transformation
and 
$\xi^\mu\equiv \epsilon \xi_1^{\mu} +\frac{1}{2}\epsilon^2\xi_2^{\mu}
+O(\epsilon^3)$.\\

Similarly the map (\ref{2.16}) generated by $\Phi$ enables us to
relate two coordinate systems $(U, x) \stackrel{\Phi}{\longrightarrow}
(U', \tilde{x})$ (see Fig.~\ref{pic1}) under an infinitesimal
transformation generated by $\epsilon \xi^{\mu}$.  In the active view
this transformation takes the point $p$ with coordinates $x^{\mu}(p)$
to the point $q = \Phi_{\epsilon}(p)$ with coordinates
$x^{\mu}(q)$. Note that in the active view it is the points that
change.
Applying the map (\ref{17a}) it follows 
\be 
\label{defcoordtrans}
{x^\mu}( {{q}})
= e^{\xi^\lambda \frac{\p}{\p x^\lambda}\big|_{{p}}} \
x^\mu( {{p}})\,, 
\ee
where we have used the fact that when acting on scalars $\pounds_{\xi}
= \xi^{\mu} \frac{\partial}{\partial_{\mu}}$ and the partial
derivatives are evaluated at $p$.
\footnote{Note, that if we had used a minus sign in the exponent of
\eq{defcoordtrans} the signs in the equation relating the coordinates
in the passive approach, \eq{passive_coord} below, would conform to
those usually found in the literature.}
The left-hand-side and the right-hand-side of \eq{defcoordtrans} are
evaluated at different points.
Equation (\ref{defcoordtrans}) can then be expanded up to second-order
as
\bea
\label{coordtrans2}
{x^\mu}(q) = x^\mu(p)+\epsilon\xi_1^{\mu}(p)
+\frac{1}{2}\epsilon^2\Big[\xi^{\mu}_{1,\nu}(p)\xi_1^{~\nu}(p)
+ \xi_2^{\mu}(p)
\Big]   \,.
\eea
Note that we do not need \eq{coordtrans2} to calculate how
perturbations change under a gauge transformation in the active
approach, it simply tells us how the coordinates of the points $p$ and
$q$ are related in this approach.

\subsubsection{Passive point of view}
\label{passive_subsect}

In the passive approach we specify the relation between two coordinate
systems directly, and then calculate the change in the metric and
matter variables when changing from one system to the other. As long
as the two coordinate systems are related through a small
perturbation, the functional form relating them is quite arbitrary.
However, in order to make contact with the active approach, discussed
above, we take \eq{coordtrans2} as our starting point.

Note, that all quantities in the passive approach are evaluated at the
same physical point. To take the passive approach further, we
therefore need to rewrite the left-hand-side and the right-hand-side
of \eq{coordtrans2}, since they are evaluated at two different
coordinate points, as described above (see also Fig.~\ref{pic2}).
We choose $p$ and $q$ to be points such, that the coordinates
of $q$ in the new coordinates are the same as the coordinates of $p$ in
the old coordinates, i.e.~$\widetilde{x^{\mu}}(q) = x^{\mu}(p)$, 
then use \eq{coordtrans2} to derive
\bea
\label{coordtrans3}
\widetilde{x^{\mu}}(q)  &=& x^{\mu}(p) \nonumber\\
&=& x^{\mu}(q) -\epsilon \xi^{\mu}_{1}(x(p)) 
-\frac{1}{2}\epsilon^{2} \Big[\xi^{\mu}_{1, \nu}(x(p))
\xi^{\nu}_{1}(x(p)) + \xi^{\mu}_{2}(x(p))\Big] \,.
\eea
Using the first terms of \eq{coordtrans2} we have
\be
x^\mu(q)= {x^\mu}(p)+\epsilon\xi_1^{\mu}(p)\,,
\ee
to get a Taylor expansion for $\xi_1^{\mu}$,
\bea
\label{xi1(p)}
\xi_1^{\mu}(p)
&=&\xi_1^{\mu}({x^\mu}(q)-\epsilon\xi_1^{\mu}(p))\nonumber\\
&=& \xi_1^{\mu}(q)-\epsilon{\xi^{\mu}_1(q)}_{,\nu}\xi_1^{~\nu}(q)\,,
\eea
where in the very last term we have replaced $\xi_1^{~\nu}(p)$ by
$\xi_1^{~\nu}(q)$, the correction being of third order.
Substituting \eq{xi1(p)} into \eq{coordtrans3} finally gives the
desired result, namely a relation between the ``old'' (untilded) and
the ``new'' (tilde) coordinate systems,
\be
\label{passive_coord}
\widetilde{x^\mu}(q)=x^\mu(q)-\epsilon\xi_1^{\mu}(q)
-\epsilon^2\frac{1}{2}\Big[
\xi_2^{\mu}(q)-{\xi^{\mu}_1(q)}_{,\nu}\xi_1^{~\nu}(q)
\Big]\,,
\ee
all evaluated at the same point $q$.

\section{Applications}
\label{appl_sect}

As an application and illustration of the above we now derive the
transformation behaviour under gauge transformations of some
quantities at first and second order. We start at first order by
highlighting the two different points of view in how the vector fields
inducing the coordinate change affect the perturbations, as detailed
above in Section \ref{gaugetrans_sect}.\\

Before studying the transformation behaviour of the perturbations, we
define and relate them to their respective backgrounds in the
following.
As our first example we choose a four-scalar, we use here the energy
density $\rho$, which can be expanded up to second order using
\eq{Texp}
\be
\rho=\rho_0+\delta\rho_1+\frac12\delta\rho_2\,,
\ee
where we already split $\delta\rho$ into its first and second order
parts according to \eq{deltaTexp}, the subscripts denoting the order
of the perturbations.

Our second example is given by the metric tensor $g_{\mu\nu}$, as
outlined in \eq{xyz}. In particular, using \eq{Texp}, the complete
Friedmann-Robertson-Walker metric tensor, up to and including
second-order perturbations, can be written as
\bea
\label{metric1}
g_{00}&=&-a^2\left(1+2\phi_1+\phi_2\right) \,, \\
g_{0i}&=&a^2\left(B_{1i}+\frac{1}{2}B_{2i}\right) \,, \\
g_{ij}&=&a^2\left[\delta_{ij}+2C_{1ij}+C_{2ij}\right]\,,
\eea
where we assumed a flat ($K=0$) background.

The first and second order perturbations $B_{1i}$ and $C_{1ij}$,
and $B_{2i}$ and $C_{2ij}$, can be further split according to
\eqs{decompBi} and (\ref{decompCij}) below into scalar, vector and tensor
parts (defined according to their transformation behaviour on spatial
3-hypersurfaces),
\bea
\label{decompBi}
B_i  &=& B_{,i} - S_i\, ,\\
\label{decompCij}
C_{ij} &=& -\psi\;\gamma_{ij} + E_{,ij}+ F_{(i,j)} + \frac{1}{2} h_{ij}\,.
\eea
where the vector parts, $S_i$ and $F_i$, are divergence free, and the
tensor part, $h_{ij}$ is divergence free and traceless, i.e.
\be
\label{divandtraceless}
S^k_{~,k}=0\,, \qquad  F^k_{~,k}=0\,, \qquad  \qquad  h^{ik}_{~,k}=0, \qquad
h^k_{~k}=0\,.
\ee
The order of the perturbations in \eqs{decompBi}, (\ref{decompCij}),
(\ref{divandtraceless}) and has been omitted in the above for ease of
presentation.
Note that $\psi$ is the curvature perturbation, describing the
intrinsic scalar curvature of spatial hypersurfaces. Furthermore
$\phi$ is the lapse function, $h_{ij}$ the tensor perturbation
describing the gravitational wave content, $B$ and $E$ describe the
scalar shear, and $S_i$ and $F_i$ the vector part of the shear.

Here and in the following we assume a flat background without loss of
generality, just simplifying our calculations and allowing us to use
partial derivatives in expressions such as \eq{decompCij}.

The perturbations are decomposed into scalar, vector, and tensor parts
since at linear order the governing equations for the different types
decouple. This is however no longer the case at higher orders, and
indeed we already see from the gauge-transformations and the
definitions of gauge-invariant variables at second order that e.g.~the
energy density (a scalar quantity) on flat hypersurfaces now also
contains first order vector and tensor parts, see \eq{rho2flat}.

Finally, we should point out that the decomposition of the metric
tensor in \eq{metric1} is not unique. This is already evident in the
temporal part of the metric tensor, where the lapse function $\phi$ is
here simply expanded in a power series,
$\phi=\phi_1+\frac{1}{2}\phi_2+\ldots$. Alternatively we could have
expanded $\exp(\phi)$ into a power series, this obviously doesn't
affect the physics.
More importantly, also the decomposition of the spatial part of the
metric tensor, that is Eq.~(\ref{decompCij}) is not unique. Indeed,
other decompositions are in use and can be just as useful or better,
depending on the circumstances and the application intended.
For example it can be useful instead of expanding $\psi$ in
\eq{decompCij} directly into a power series, to expand $e^\psi$ (see
e.g.~Ref.~\cite{SB}, and for a relation of the two expansions
Ref.~\cite{LMS}).

\subsection{Passive point of view}
\label{passive_appl_sect}

The passive point of view is very popular at first order, see e.g.~the
original paper by Bardeen \cite{Bardeen80}, the review by Kodama and
Sasaki \cite{KS}, and the one by Mukhanov, Feldman, and Brandenberger
\cite{MFB}.

The starting point in the {passive approach} is to identify an
invariant quantity, that allows to relate quantities to be evaluated
in the two coordinate systems. We denote the two coordinate systems by
$\tilde x^\mu$ and $x^\mu$ system, and their relation is given by
\eq{passive_coord}.
We choose as an example the energy density, $\rho$, which as
a four scalar won't change (however, once it has been split into
different orders, it will change).
Another invariant is the line element $ds^2$, which allows to study
the transformation properties of the metric tensor, by exploiting the
invariance of $ds^2$, i.e.,
\be
ds^2=\tilde g_{\mu\nu}d \tilde x^\mu d\tilde x^\nu
= g_{\mu\nu}dx^\mu d x^\nu\,,
\ee
which we here will not pursue, but see e.g.~\cite{KS,thesis}.

Turning instead to the energy density as an illustrative example, we
get the transformation behaviour of the perturbation from the
requirement that it has to invariant under a change of coordinate
system and therefore has to be the same in the $\tilde x^\mu$ and the
$x^\mu$ system, that is
\bea
\label{tilderho}
\tilde\rho(\tilde x^\mu)=\rho ( x^\mu)\,.
\eea
To first order, the two coordinate systems are related, using the
linear part of \eq{passive_coord}, by 
\be 
\tilde x^\mu=x^\mu-\xi_1^\mu\,.  
\ee
Before we can study the transformation behaviour of the
perturbations at first order, we split the generating vector
$\xi_1^\mu$ into a scalar temporal part $\alpha_1$ and a spatial
scalar and vector part, $\beta_{1}$ and $\gami1$, according to
\be
\label{def_xi1}
\xi_1^\mu=\left(\alpha_1,\beta_{1,}^{~~i}+\gami1\right)\,,
\ee
where the vector part is divergence-free $\p_k\gamk1=0$.
Then expanding the left-hand side of \eq{tilderho}, we get neglecting
terms of $O(\epsilon^2)$,
\bea
\tilde\rho(\tilde x^\mu)&=&\tilde\rho(x^\mu-\xi^\mu)\nonumber\\
&=&\tilde\rho ( x^\mu) - \tilde\rho_{,\lambda}\xi^\lambda\nonumber\\
&=&\tilde\rho_0+\tilde\delta\rho_1-\rho_0'\alpha_1
\,,
\eea
and similarly expanding the right-hand side of \eq{tilderho}, we have
\be
\rho(x^\mu)=\rho_0(x^\mu)+\delta\rho_1(x^\mu)\,.
\ee
Finally, since by assumption $\tilde\rho_0(x^\mu)=\rho_0(x^\mu)$, we
get
\be
\label{rho_trans_passive1}
\tilde\delta\rho_1=\delta\rho_1+\rho_0'\alpha_1\,.
\ee
Note that all quantities are evaluated at the same physical point.

\subsection{Active point of view}
\label{active_appl_sect}

We now turn to the active point of view when calculating the effect of
gauge transformations on perturbations. Here, as detailed in Section
\ref{gaugetrans_sect} above, one actively maps the perturbed
quantities from one manifold to another. The relation of the
coordinate systems on the two manifolds is also induced by the map.

At first order the preference of which approach to use is a question
of taste, and as pointed out above most first order papers use passive
view point, but see e.g.~\cite{MW2004} for first order active
calculation. However, at second order we found the active view point
easier to implement, and it is used in many other second order works,
e.g.~Refs.~\cite{Mukhanov96,bruni,Noh2004}.

\subsubsection{First order}

As in the passive view section above, we start with the energy
density.
It follows immediately from \eq{2.18} that to first order a scalar
quantity such as the energy density transforms as
\be
\label{transrho_1}
\widetilde\delta\rho_1 = \delta\rho_1 + \rho_0'\alpha_1 \,.
\ee

The transformations of the first order metric perturbations also
follow from \eq{2.18}. 
We then find that the metric tensor transforms at first order, as
\bea
\label{general_gmunu1}
\wt{\delta g^{(1)}_{\mu\nu}}&=&\delta g^{(1)}_{\mu\nu}
+g^{(0)}_{\mu\nu,\lambda}\xi^\lambda_1
+g^{(0)}_{\mu\lambda}\xi^\lambda_{1~,\nu}
+g^{(0)}_{\lambda\nu}\xi^\lambda_{1~,\mu}
\eea

As another example we now turn to the spatial part of the metric
tensor.
Note that \eq{2.18} gives only the transformation of the total spatial
part of the metric, $C_{ij}$. If we then ask how the components of
$C_{ij}$ transform, we have to use \eq{divandtraceless}.

To get the change of the metric functions in the spatial part of
the metric under a gauge transformation, we get the transformation of the
spatial part of the metric ${\delta g^{(1)}_{ij}}$, and hence $C_{1ij}$, 
from \eq{general_gmunu1} as
\bea
\label{Cij1trans}
2\wt C_{1ij}=2 C_{1ij} +2\H\alpha_1\delta_{ij}+\xi_{1i,j}+\xi_{1j,i}\,,
\eea
where we reproduce \eq{decompCij} above for convenience at first order,
\be
\label{decompCij1}
2 C_{1ij}=-2\psi_1\delta_{ij}+2E_{1,ij}+2 F_{1(i,j)}+h_{1ij}\,.
\ee
Taking the trace of \eq{Cij1trans} and substituting in \eq{decompCij1} we get
\bea
\label{Cij1trace}
-3\wt\psi_1+\nabla^2\wt E_1
=-3\psi_1+\nabla^2 E_1+3\H\alpha_1+\nabla^2\beta_{1}\,.
\eea
Now applying the operator $\p^i\p^j$ to \eq{Cij1trans} we get a second
equation relating the scalar perturbation $\psi_1$ and $E_1$,
\bea
\label{Cij1divdiv}
-3\wt\nabla^2\psi_1+\nabla^2\nabla^2\wt E_1
=-3\nabla^2\psi_1+\nabla^2\nabla^2 E_1+3\H\nabla^2\alpha_1
+\nabla^2\nabla^2\beta_1\,. 
\eea
Taking the divergence of \eq{Cij1trans} we get 
\bea
2\wt C_{1ij,}^{~~~j}=2 C_{1ij,}^{~~~j} +2\H\alpha_{1,i}
+\nabla^2\xi_{1i}+\nabla^2\beta_{1,i}\,. 
\eea
Substituting in our results for $\wt\psi_1$ and $\wt E_1$ we the
arrive at
\be
\nabla^2 \wt F_{1i}= \nabla^2 F_{1i}+\nabla^2 \gami1\,.
\ee

We can sum up the well known transformations of the first order metric
perturbations we have from the above, first for the scalars as
(e.g.\cite{MW2004})
\bea
\label{transphi1}
\widetilde {\A1} &=& \A1 +\H\alpha_1+\alpha_1'\,,\\
\label{transpsi1}
\widetilde \psi_1 &=& \psi_1-\H\alpha_1 \,,\\
\label{transB1}
\widetilde B_1 &=& B_1-\alpha_1+\beta_1'\,,\\
\label{transE1}
\widetilde E_1 &=& E_1+\beta_1\,,
\eea
where $\H = a'/a$, and for the vector perturbations as
\bea
\label{transS1}
\widetilde {S_{1}^{~i}} &=& S_{1}^{~i}-\gami1'\,, \\
\label{transF1}
\widetilde {F_{1}^{~i}} &=& F_{1}^{~i}+\gami1\,. 
\eea

The first order tensor perturbation is found to be gauge-invariant,
\be
\widetilde h_{1ij} = h_{1ij}\,.
\ee
by substituting \eqs{transphi1} to (\ref{transF1}) into
\eq{Cij1trans}.
This can also be understood from the Stewart-Walker lemma
\cite{walker}: at first order, quantities that are identically zero in
the background are manifestly gauge-invariant, and there is no tensor
part in the background. However, as we shall see below, this only works
for quantities at the next higher order: for example the second order
tensor perturbations will in general not be gauge-invariant.

\subsubsection{Constructing gauge-invariant variables at first order}
\label{construct_sect1}

To construct a gauge-invariant quantity, say the energy density on
flat slices, that is hypersurfaces on which $\widetilde \psi_1=0$, we
see from \eq{transpsi1} that this gives
\be
\label{flat_def1}
\alpha_1 =\frac{\psi_1}{\H}\,.
\ee
All we need to do next is to substitute \eq{flat_def1} into
\eq{transrho_1}, and get a gauge-invariant in the sense of being
independent of gauge artifacts,for example the energy density on
flat slices
\be
\label{example1}
\delta\rho_1\Big|_{\rm{flat}} = \delta\rho_1
+\frac{\rho_0'}{\H} \psi_1\,.
\ee
This is gauge-invariant in the $\xi^\mu$-independence sense, but it does
depend on the choice of background (e.g.~a background depending on
$x^i$ instead of just time as in FRW would obviously give a very
different result). This works for all the perturbations and also
at second order and higher.

We conclude the above example by observing that to remove the gauge
modes on sub-horizon scales, often referred to as specifying the
\emph{threading}, we can choose $\widetilde E_1=0$, which gives
\be
\label{flat_def1a}
\beta_1 =-E_1\,.
\ee
For the vector modes we choose 
$\widetilde F_1^i=0$, which gives
\be
\label{flat_def1b}
\gami1 =-F_1^i\,.
\ee
Hence in this gauge the spatial part of the perturbed metric is zero
with the exemption of the tensor-modes.

\subsubsection{Second order}

At second order the generating vector $\xi_2^\mu$ is split into
a scalar time and scalar and vector spatial part, similarly as at
first order, 
\be
\label{def_xi2}
\xi_2^\mu=\left(\alpha_2,\beta_{2,}^{~~i}+\gami2\right)\,,
\ee
where the vector part is divergence-free $\p_k\gamk2=0$.
We then find from Eqs.~(\ref{2.18}) that a four scalar transforms 
at second order
\bea
\label{rhotransform2}
\widetilde\delta\rho_2 = \delta\rho_2
+\rho_0'\alpha_2&+&\alpha_1\left(
\rho_0''\alpha_1+\rho_0'{\alpha_1}'+2\delta\rho_1'\right)\nonumber\\
&+&\left(2\delta\rho_1+\rho_0'{\alpha_1}\right)_{,k} 
(\beta_{1,}^{~~k}+\gamk1)
\,.
\eea
We see here already the coupling between vector and scalar
perturbations in the last term through the gradient and $\gami1$.  The
gauge is only specified once the scalar temporal gauge perturbations
at first and second order, $\alpha_1$ and $\alpha_2$, and the first
order spatial gauge perturbations, $\beta_{1}$ and $\gami1$, are
specified.\\

The metric tensor transforms at second order, from \eq{2.18} 
as
\bea
\label{general_gmunu2}
\wt{\delta g^{(2)}_{\mu\nu}}&=&\delta g^{(2)}_{\mu\nu}
+g^{(0)}_{\mu\nu,\lambda}\xi^\lambda_2
+g^{(0)}_{\mu\lambda}\xi^\lambda_{2~,\nu}
+g^{(0)}_{\lambda\nu}\xi^\lambda_{2~,\mu}
+2\Big[
\delta g^{(1)}_{\mu\nu,\lambda}\xi^\lambda_1
+\delta g^{(1)}_{\mu\lambda}\xi^\lambda_{1~,\nu}
+\delta g^{(1)}_{\lambda\nu}\xi^\lambda_{1~,\mu}
\Big]\nonumber \\
&&+g^{(0)}_{\mu\nu,\lambda\alpha}\xi^\lambda_1\xi^\alpha_1
+g^{(0)}_{\mu\nu,\lambda}\xi^\lambda_{1~,\alpha}\xi^\alpha_1
+2\Big[
g^{(0)}_{\mu\lambda,\alpha} \xi^\alpha_1\xi^\lambda_{1~,\nu}
+g^{(0)}_{\lambda\nu,\alpha} \xi^\alpha_1\xi^\lambda_{1~,\mu}
+g^{(0)}_{\lambda\alpha}  \xi^\lambda_{1~,\mu} \xi^\alpha_{1~,\nu}
\Big]
\nonumber \\
&&+g^{(0)}_{\mu\lambda}\left(
\xi^\lambda_{1~,\nu\alpha}\xi^\alpha_1
+\xi^\lambda_{1~,\alpha}\xi^\alpha_{1,~\nu}
\right)
+g^{(0)}_{\lambda\nu}\left(
\xi^\lambda_{1~,\mu\alpha}\xi^\alpha_1
+\xi^\lambda_{1~,\alpha}\xi^\alpha_{1,~\mu}
\right)\,.
\eea
%
%
Now following similar lines as at first order in the previous section,
we could get the transformation behaviour for the second order lapse
function $\phi_2$ straight from the $0-0$-component of
\eq{general_gmunu2}.

Instead, to keep the discussion as brief as possible, we now turn to
the transformation behaviour of the perturbations in the spatial part
of the metric tensor. Here we can follow a similar procedure as in the
linear case. But, the task is made more complicated not only by the
size of the expressions but more importantly by the fact that now we
will have to let inverse Laplacians operate on products, in order to
get the transformations of the scalar, vector, and tensor parts of the
spatial metric.

Using \eq{general_gmunu2} we find that the perturbed spatial part of
the metric, $C_{2ij}$, transforms at second order as
\bea
\label{Cij2trans}
2\widetilde C_{2ij}&=&2C_{2ij}+2\H\alpha_2 \delta_{ij}
+\xi_{2i,j}+\xi_{2j,i}+\X_{ij}\,,
\eea
where we defined $\X_{ij}$ to contain the terms quadratic in the first
order perturbations as
\bea
\label{Xijdef}
\X_{ij}&\equiv&
2\Big[\left(\H^2+\frac{a''}{a}\right)\alpha_1^2 
+\H\left(\alpha_1\alpha_1'+\alpha_{1,k}\xi_{1}^{~k}
\right)\Big] \delta_{ij}\nonumber\\
&&
+4\Big[\alpha_1\left(C_{1ij}'+2\H C_{1ij}\right)
+C_{1ij,k}\xi_{1}^{~k}+C_{1ik}\xi_{1~~,j}^{~k}
+C_{1kj}\xi_{1~~,i}^{~k}\Big]
+2\left(B_{1i}\alpha_{1,j}+B_{1j}\alpha_{1,i}\right)
\nonumber\\
&&
+4\H\alpha_1\left( \xi_{1i,j}+\xi_{1j,i}\right)
-2\alpha_{1,i}\alpha_{1,j}+2\xi_{1k,i}\xi_{1~~,j}^{~k}
+\alpha_1\left( \xi_{1i,j}'+\xi_{1j,i}' \right)
+\left(\xi_{1i,jk}+\xi_{1j,ik}\right)\xi_{1}^{~k}
\nonumber\\
&&+\xi_{1i,k}\xi_{1~~,j}^{~k}+\xi_{1j,k}\xi_{1~~,i}^{~k}
+\xi_{1i}'\alpha_{1,j}+\xi_{1j}'\alpha_{1,i}
\,.
\eea
Note that in \eq{Xijdef} above and in the following we will not decompose the
spatial part of \eq{def_xi1}, $\xi_1^i=\beta_{1,}^{~~i}+\gami1$,
whenever convenient to keep the presentation as compact as possible.

The perturbed spatial part of the metric, $C_{2ij}$, is decomposed in
\eq{decompCij} above into scalar, vector, and tensor part, which we
reproduce here at second order,
\be
\label{decompCij2}
2 C_{2ij}=-2\psi_2\delta_{ij}+2E_{2,ij}+2 F_{2(i,j)}+h_{2ij}\,.
\ee
Taking the trace of \eq{Cij2trans} and substituting in \eq{decompCij2} we get
\bea
\label{Cij2trace}
-3\wt{\psi_2}+\nabla^2\wt{E_2}
&=&-3{\psi_2}+\nabla^2 {E_2}
+3\H\alpha_2 +\nabla^2\beta_2
+\frac{1}{2}\X^k_{~k}\,,
\eea
where we find $\X^k_{~k}$ to be 
\bea
\label{Xtrace}
\frac{1}{2}\X^k_{~k}&=&
3\left(\H^2+\frac{a''}{a}\right)\alpha_1^2
+3\H\left(\alpha_1\alpha_1'+\alpha_{1,k}\xi_{1}^{~k}
\right)\nonumber\\
&&
+2\Big[\alpha_1\left( {C_{1~~k}^{~k}}'+2\H{C_{1~~k}^{~k}}
\right)
+ {C_{1~~k,l}^{~k}}\xi_{1}^{~l}
+2C_{1}^{~kl}\xi_{1l,k}\Big]
+2B_{1k}\alpha_{1,}^{~~k}\\
%
&&
-\alpha_{1,k}\alpha_{1,}^{~~k}
+2\xi_{1~,}^{k~~l}\xi_{1(k,l)}
+\alpha_1\nabla^2\left(\beta_1'+4\H\beta_1\right)
+\nabla^2\beta_{1,k}\xi_{1}^{~k}
+\xi_{1k}'\alpha_{1,}^{~~k}\nonumber
\,.
\eea
Now applying the operator $\p^i\p^j$ to \eq{Cij2trans} we get a second
equation relating the scalar perturbations $\psi_2$ and $E_2$,
\be
\label{Cij2divdiv}
-\wt\nabla^2\psi_2+\nabla^2\nabla^2\wt E_2
=-\nabla^2\psi_2+\nabla^2\nabla^2 E_2+\H\nabla^2\alpha_2
+\nabla^2\nabla^2\beta_2 +\frac{1}{2} \X^{ij}_{~~,ij}  \,,
\ee
This gives for the transformations of the curvature perturbation at
second order,
\be
\label{transpsi2}
\wt\psi_2=\psi_2-\H\alpha_2-\frac{1}{4}\X^k_{~k}
+\frac{1}{4}\nabla^{-2} \X^{ij}_{~~,ij}\,,
\ee
and for the shear scalar,
\be
\label{transE2}
\wt E_2=E_2+\beta_2+\frac{3}{4}\nabla^{-2}\nabla^{-2}\X^{ij}_{~~,ij}
-\frac{1}{4}\nabla^{-2}\X^k_{~k}\,.
\ee

Taking the divergence of \eq{Cij2trans} we get 
\be
2\wt C_{2ij,}^{~~~j}=2 C_{2ij,}^{~~~j} +2\H\alpha_{2,i}
+\nabla^2\xi_{2i}+\nabla^2\beta_{2,i}+\X_{ik,}^{~~k}\,. 
\ee
Substituting in our results for $\wt\psi_2$ and $\wt E_2$ we then
arrive at
\be
\nabla^2 \wt F_{2i}= \nabla^2 F_{2i}+\nabla^2 \gamma_{2i}
+\X_{ik,}^{~~~k}-\nabla^{-2}\X^{kl}_{~~,kli}
\,.
\ee
Finally
\be
\wt F_{2i}= F_{2i}+\gamma_{2i}
+\nabla^{-2}\X_{ik,}^{~~~k}-\nabla^{-2}\nabla^{-2}\X^{kl}_{~~,kli}
\,.
\ee

We finally turn to the tensor perturbation at second
order. Substituting our previous results for $\psi_2$, $E_2$, and
$F_{2i}$ into \eq{Cij2trans} we get, probably surprisingly,
\bea
\label{transhij2}
\wt h_{2ij}&=& h_{2ij}+\X_{ij}
+\frac{1}{2}\left(\nabla^{-2}\X^{kl}_{~~,kl}-\X^k_{~k}
\right)\delta_{ij}
+\frac{1}{2}\nabla^{-2}\nabla^{-2}\X^{kl}_{~~,klij}\nonumber\\
&&+\frac{1}{2}\nabla^{-2}\X^k_{~k,ij}
-\nabla^{-2}\left(\X_{ik,~~~j}^{~~~k}+\X_{jk,~~~i}^{~~~k}
\right)
\,.
\eea
Although the second-order tensor transformation $h_{2ij}$ is not
dependent on the second-order gauge-functions $\xi_{2}^\mu$, it does
depend on first order quantities quadratically.

The same holds for other quantities that are zero in the background:
the first order quantity is gauge-invariant by virtue of the
Stewart-Walker lemma \cite{walker} (and by construction). However the
second order quantity is no longer gauge-invariant, as shown above in
the case of the tensor perturbation, $h_{2ij}$. This is not a violation
of the Stewart-Walker lemma, it merely shows that the second order
quantities ``live'' in a first order ``background''. 
Another example is the anisotropic stress, which is gauge-invariant at
first order, but not at second.

\subsubsection{Constructing gauge-invariant variables at second order}
\label{construct_sect2}

We can now construct, just as at first order in Section
\ref{construct_sect1}, gauge-invariant variables at second order. We
choose the same example as in the previous section, namely the energy
density on flat hypersurfaces. But now also give the second order
tensor perturbation in this gauge.

To specify the gauge at second order we choose hypersurfaces on which
$\widetilde \psi_2=0$, and we see from \eq{transpsi2} that this gives
\bea
\label{flat_def2}
\alpha_{2\fg}=\frac{\psi_2}{\H}+\frac{1}{4\H}\left[
\nabla^{-2}\X^{ij}_{\fg,ij}-\X^k_{\fg k}\right]\,,
\eea
where we get $\X_{\fg ij}$ from \eq{Xijdef} using the first order gauge
generators given above, as
\bea
\label{Xijflat}
\X_{\fg ij}&=& 2\left[
\psi_1\left(\frac{\psi_1'}{\H}+2\psi_1\right)+\psi_{1,k}\xi_{1\fg}^k
\right]\delta_{ij}
+\frac{4}{\H}\psi_1\left(C_{1ij}'+2\H C_{1ij}\right)\nonumber\\
&&
+4 C_{1ij,k}\xi_{1\fg}^k
+\left(4 C_{1ik}+\xi_{1\fg i,k}\right)\xi_{1\fg,j}^k
+\left(4 C_{1jk}+\xi_{1\fg j,k}\right)\xi_{1\fg,i}^k\nonumber\\
&&
+\frac{1}{\H}\Big[
\psi_{1,i}\left(2B_{1j}+\xi_{1\fg j}'\right)
+\psi_{1,j}\left(2B_{1i}+\xi_{1\fg i}'\right)
\Big]
-\frac{2}{\H^2}\psi_{1,i}\psi_{1,j}\nonumber\\
&&
+\frac{2}{\H}\psi_1\left(
\xi_{1\fg (i,j)}'+4\H  \xi_{1\fg (i,j)}\right)
+2\xi_{1\fg}^k\xi_{1\fg (i,j)k}
+2\xi_{1\fg k,i}\xi_{1\fg,j}^k\,,\nonumber\\
\eea
where we defined
\be
\xi_{1\fg i}=-\left(E_{1,i}+F_{1i}\right)\,.
\ee
The trace of \eq{Xijflat} is then
\bea
\X^k_{\fg k}&=&6\left[
\psi_1\left(\frac{\psi_1'}{\H}+2\psi_1\right)+\psi_{1,k}\xi_{1\fg}^k
\right]
+\frac{4}{\H}\psi_1\left(C_{1~k}^{k\prime}+2\H C_{1~k}^{k} \right)\nonumber\\
&&
+4 C_{1~k,l}^{k}\xi_{1\fg}^l
+4\left(2 C_{1}^{kl}+\xi_{1\fg ,}^{k~~~~l}\right)\xi_{1\fg (k,l)}
-2\nabla^2E_{1,k}\xi_{1\fg}^k
\\
&&
+\frac{2}{\H}\left(
2B_{1k}+\xi_{1\fg k}'-\frac{1}{\H}\psi_{1,k}\right)\psi_{1,}^{~k}
-\frac{2}{\H}\left(\psi_1\nabla^2E_1'+4\H\nabla^2E_1\right)
\,.\nonumber
\eea

Then substituting at first order \eqs{flat_def1}, (\ref{flat_def1a}),
and (\ref{flat_def1b}), and at second order \eq{flat_def2} into
\eq{rhotransform2}, we get the second order energy density
perturbation on uniform curvature hypersurfaces \cite{MW2003}
\bea
\label{rho2flat}
\widetilde{\delta\rho_{2\fg}} &=&\delta\rho_2
+\frac{\rho_0'}{\H}\psi_2
+\frac{\rho_0'}{4\H}\left(
\nabla^{-2}\X^{ij}_{\fg,ij}-\X^k_{\fg k}\right)
\\
&&+\frac{\psi_1}{\H^2}\Big[\rho_0'' {\psi_1}
+\rho_0'\left(\psi_1'-\frac{\H'}{\H}\psi_1\right)+2\H\delta\rho_1'\Big]
+\left(2\delta\rho_1+\frac{\rho_0'}{\H}\psi_1\right)_{,k}\xi_{1\fg}^k
\,.\nonumber
\eea

The second order tensor perturbation in that the flat gauge, i.e.~on
uniform curvature hypersurfaces, is given by substituting
\eqs{flat_def1}, (\ref{flat_def1a}), and (\ref{flat_def1b}), into
\eq{transhij2}, and we find after some algebra
\bea
\label{hij2flat}
\wt h_{2\fg ij}&=& h_{2ij}+\X_{\fg ij}
+\frac{1}{2}\left(\nabla^{-2}\X^{kl}_{\fg,kl}-\X^k_{\fg k}
\right)\delta_{ij}
+\frac{1}{2}\nabla^{-2}\nabla^{-2}\X^{kl}_{\fg ,klij}\nonumber\\
&&+\frac{1}{2}\nabla^{-2}\X^k_{\fg k,ij}
-\nabla^{-2}\left(\X_{\fg ik,j}^{~~~k}+\X_{\fg jk,i}^{~~~k}
\right)
\,.
\eea

\section{Discussion and conclusions}
\label{dissc_sect}

This is neither the first, nor will it be the last, discussion of
perturbation theory in cosmology. However, in this concise
introduction we have tried to strike a balance between mathematical
rigour and ease of application of the results. For a more detailed
exposition of cosmological perturbation theory and further references
see Ref.~\cite{MW2008}.

We have here studied perturbations about a flat FRW background
spacetime. But as pointed out above, the formalism introduced by
Bardeen can easily be applied to other settings and background
spacetimes, and can also be extended beyond GR. Indeed, perturbation
theory and the formalism discussed in this paper can be applied to all
covariant metric theories.
Although here we have assumed standard four dimensional (4D) Einstein
gravity throughout, the formalism has also been applied, for
example, to 5D braneworld models (see e.g.~Ref.~\cite{roy} for an
overview), and has been used to construct gauge-invariant variables in
that theory.

%
Whereas most of the material discussed in the previous sections has
been expounded elsewhere, albeit often in different form and with
other aims, we are not aware of the derivation of how the second order
tensor perturbations transforms in full generality under
gauge-transformations being discussed elsewhere (see however
Ref.~\cite{pitrou} for the case of scalar perturbations). Also its
representation in the uniform curvature gauge has been discussed for
the first time. These results will be of particular interest in second
order calculations of the gravitational wave background
\cite{kishore}.
The transformations at second order of the decomposed components of
the spatial parts of the metric have also not been discussed in the
literature before, and will be particularly useful in relating
quantities calculated in different gauges.

\acknowledgments

The authors are grateful to Adam Christopherson, David Seery, and David
Wands for useful discussions and comments.

\appendix
\section{Definitions}

In this appendix we bring together some key ideas and definitions from
differential geometry, in particular those regarding maps of
manifolds, which are useful in setting up perturbation theory in
general relativity where covariance matters. This appendix is not a
comprehensive study as our intention is only to show, intuitively, why
certain expressions take the form they do and to define some key
ideas. In order to be useful to those who do not wish to enter into
the formalism of differential geometry we will use coordinate
expressions where possible. For those who worry about these things we
will therefore be working in coordinate neighbourhoods and all
functions will be assumed to be adequately differentiable as they are
for most of cosmology. For more details and a less coordinate
dependent approach the reader is referred to the books by Hawking and
Ellis \cite{hande} and by Wald \cite{wald}. \\

\begin{itemize}

\item {\bf Maps between Manifolds} \\
We assume some familiarity with the definition and properties of
differential manifolds and are concerned here with maps between such
manifolds, in particular a diffeomorphism.  Consider two manifolds
${\cal M}$ and ${\cal M_{\eps}}$ and denote the chart maps
(coordinates) on each by
\begin{eqnarray}
 f_{a}: O_{a} & \rightarrow & U_{a} \ \ \ {\rm for} \ \ \ O_{a}
 \subset {\cal M}, \, U_{a} \subset \RR^{n} \,, \nonumber \\
 g_{b}: O_{b} & \rightarrow & U_{b} \ \ \ {\rm for} \ \ \ O_{b}
 \subset {\cal M_{\eps}}, \, U_{b} \subset \RR^{n}\,.
 \end{eqnarray}
In words the function $f_{a}$ assigns coordinates (in $U_{a}$) to
points in the n-dimensional neighbourhood $O_{a}$ of ${\mathcal
M}$ and $g_{b}$ does the same in the neighbourhood $O_{b}$ of $\Me$.

The map
\be
\phi: {\cal M} \rightarrow {\cal M_{\eps}}\,,
\ee
is $C^{\infty}$, i.e. infinitely continuously differentiable in the
 advanced calculus sense, if for each $a$ and $b$ the map
\be
\label{map}
f_{a} \circ \phi^{-1} \circ g^{-1}_{b}: U_{b} \rightarrow U_{a}\,,
\ee
is $C^{\infty}$. The map $\phi$ is a diffeomorphism if it is
one-to-one, onto and $\phi$ and its inverse $\phi^{-1}$ are
$C^{\infty}$. Loosely speaking, for coordinates \{$x^{i}$ on $U_{a}$
and $y^{i}$ on $U_{b}$ \}, \eq{map} relating the coordinates under the
map $\phi$ can be represented by the equations
\be
y^{\mu} = \phi^{\mu}\,(x^{\nu})\,,
\ee
with inverse
\be
x^{\mu} = (\phi^{-1})^{\mu}\,(y^{\nu})\,,
\ee
where $\phi^{\mu}$ and $(\phi^{-1})^{\mu}$ are $C^{\infty}$.

\item {\bf Maps of Vectors induced by mapping manifolds}

A manifold map $\phi: {\cal M} \rightarrow {\cal M_{\eps}}$
induces a map of the tangent vectors at $p$ on ${\cal M}$ to
tangent vectors at $\phi(p)$ on ${\cal M}_{\eps}$.  We will write
this map as $\phi^{*}: V_{p} \rightarrow V_{\phi(p)}$ where
$V_{p}$ and $V_{\phi(p)}$ denote the tangent spaces at $p$ and
$\phi(p)$. Using the coordinate description above, the map
$\phi^{*}$ can be written as
\be
(\phi^{*})^{\mu}_{\nu} = \frac{\partial \phi^{\mu}}{\partial x^{\nu}}\,,
\ee
and if $X^{\mu} \in V_{p}$ and $Y^{\nu} \in V_{\phi(p)}$ then
\be
\left(\phi^{*}\right)^{\mu}_{\nu}:\, X^{\nu} \rightarrow Y^{\mu} \,,
\ \ \ {\rm or} \ \ \ 
Y^{\mu} \, = \, (\phi^{*})^{\mu}_{\nu}\; 
X^{\nu}\; = \frac{\partial
\phi^{\mu}}{\partial x^{\nu}}\; X^{\nu}\,,
\ee
in the usual notation.  The map $\phi^{*}$ is sometimes called a
pushforward.

\item{\bf Pullback}

The concept of a pullback is well described by its name. If we are
 given a function $f$ defined on a manifold ${\cal N}$ so
\be
f: {\cal N} \rightarrow \RR^{n}\,,
\ee
and a manifold map
\be
\phi: {\cal M} \rightarrow {\cal N}\,,
\ee
then $\phi$ can be used to {\it pullback} $f$ from ${\cal N}$ to
${\cal M}$ by using the composite map $f \circ \phi$. This pulls $f$
back from ${\cal N}$ to map ${\cal M}$ to $\RR^{n}$, i.e.  
\be
f \circ \phi: {\cal M} \rightarrow \RR^{n} \,.
\ee
This is perhaps easier to follow if viewed as a sequence: the $\phi$
maps ${\cal M}$ to ${\cal N}$ then $f$ maps ${\cal N}$ to $\RR^{n}$
or, in symbols, 
\be
f \circ \phi: {\cal M} \longrightarrow f : {\cal N} \longrightarrow
\RR^{n}\,.
\ee
The map $f \circ \phi$ is the pullback of $f$.

\item {\bf Maps of covectors induced by mapping manifolds}

For covectors (forms or covariant vectors) the map between manifolds
$\phi: {\cal M} \rightarrow {\cal M}_{\eps} $ induces a pullback map
$\phi_{*}$ which takes covectors at $\phi (p)$ on ${\cal M}_{\eps}$ to
covectors at $p$ on ${\cal M}$, i.e.,
\be
\phi_{*}\, : \, W_{\phi(p)}\; \rightarrow \; W_{p}\,,
\ee
where $W_{\phi(p)}$ denotes the cotangent space at $\phi(p)$ on ${\cal
M}_{\eps}$ and $W_{p}$ is the cotangent space at $p$ on ${\cal M}$.
This can be written in coordinates, by letting $X_{\mu} \in W_{p}$ and $Y_{\mu}
\in W_{\phi(p)}$, and we can write
\be
X_{\mu} = (\phi_{*})^{\nu}_{\mu}\, Y_{\nu} = \frac{\partial \phi^{\nu}}
{\partial x^{\mu}} \, Y_{\nu} \,, \qquad
%
\mathrm{or} \qquad
%
Y_{\nu} 
= \left(\frac{\partial \phi^{\nu}}{\partial x^{\mu}}\right)^{-1} 
\, X_{\mu} \,.
\ee
The mapping of covariant and contravariant tensors of higher rank
follows the same pattern as that for the vectors and covectors.
Tensors with mixed co- and contravariant indices also follow this
pattern, although proving it requires a little care (see \cite{wald}
p.~438). A pullback on a vector field effectively reverses the effect
of a pushforward and so the definitions of pullback and pushforward
act to preserve the scalar property of the product of a vector and a
covector.

\item {\bf Pullback of a Composite Map}

Note that for composites of maps the pullback behaves in a
different way to the push forward.  For pushforward maps,
\be
\phi: {\cal M}  \rightarrow  {\cal N} \,, \qquad
%
\mathrm{and} \qquad
%
\psi: {\cal N} \rightarrow  {\cal P} \,,
\ee
the composite is simply
\be
\psi \circ \phi: {\cal M}  \rightarrow  {\cal P} \,.
\ee
For the corresponding pullback maps we have
\begin{eqnarray}
\label{A6}
\phi_{*}: {\cal N} & \rightarrow & {\cal M} \,,
\qquad
\psi_{*}: {\cal P}   \rightarrow  {\cal N} \,, \\
(\psi \circ \phi)_{*}: {P} & \rightarrow & {\cal M} \,,
\end{eqnarray}
but, and this is important, if we remove the bracket we have to
reverse the maps
\be
\phi_{*} \circ \psi_{*}: {\cal P} \rightarrow  {\cal M} \,,
\label{A9}
\ee
i.e.~$(\psi \circ \phi)_{*} = \phi_{*} \circ \psi_{*}$\, as can be
seen by looking at the actions of the maps themselves.

\item {\bf Lie Derivative}

Let ${\cal M}$ be a differential manifold and let ${\bf \xi}$ be a
vector field on ${\cal M}$, then ${\bf \xi}$ generates a map of
${\cal M}$ onto itself as follows. In a coordinate neighbourhood
solve the system of ordinary differential equations
\be
\frac{d x^{\mu}}{d \epsilon} = \xi^{\mu} \,,
\ee
in ${\mathbb{R}}^{n}$. Given initial points in ${\mathcal M}$ at $\epsilon
= 0$ this will generate a unique one parameter family of integral
curves with one and only one curve through each point in a
neighbourhood.  For an integral curve $\gamma$ let
$\phi_{\eps}(p)$ be the point a distance $\eps$ along $\gamma$
from $p$. Applied to the family of curves $\phi_{\eps}$ generates
a one parameter family of diffeomorphisms of ${\mathcal M}$ onto
itself.  Clearly $\phi_{\eps_{1}}\circ\phi_{\eps_{2}} =
\phi_{\eps_{1}} + \phi_{\eps_{2}}$ and $\phi_{0}$ is the identity
map.

\noindent Now let $T$ be a tensor field on ${\cal M}$ then the
pullback $\phi_{*}$ of $\phi$ defines a new tensor field $\phi_{*
\eps}T$ on ${\cal M}$ which is a function of $\eps$. This enables
us to define the Lie Derivative, a covariant differentiation on
${\cal M}$, which does not increase the rank of the tensor.
\be
\pounds_{\xi} T := \lim_{\eps \rightarrow
0}\frac{1}{\eps}\left(\phi_{* \eps}\, T - T \right) \,, \label{A10}
\ee
This can be interpreted as follows - the map $\phi_{* \eps}$ pulls
back the value of $T$ at $\phi_{\eps}(p)$ to $p$ from which we
subtract the actual value of $T$ at $p$.  This difference is a well
defined tensor quantity since the difference is taken at the point
$p$. We can now take the limit to obtain a meaningful derivative at
$p$ - called the Lie Derivative.

\item {\bf Exponential operator}

Given an operator $A$ the exponential operator is defined by the
formal series
\be 
e^{A} := 1 +  A + \frac{1}{2} A^{2} + \sum_{n = 3}^{\infty}
\left(\frac{1}{n!}\right) A^{n}  \,.
\ee
Thus, for instance, if $A = \eps \pounds_{\xi}$ is the Lie
Derivative operator multiplied by $\eps$ the corresponding
exponential operator is given by
\be
e^{\eps \pounds_{\xi}} := 1 + \eps \pounds_{\xi}
+ \frac{\eps^{2}}{2}\pounds_{\xi}^{2}
+ \sum_{n = 3}^{\infty} \left(\frac{\eps^{n}}{n!}
   \right) \pounds_{\xi}^{n} \,.
\ee
\end{itemize}



\end{document}